\documentclass[10pt,letterpaper,twocolumn]{article}
\usepackage{OL}

\long\def\symbolfootnote[#1]#2{\begingroup%
\def\thefootnote{\fnsymbol{footnote}}\footnote[#1]{#2}\endgroup}

\begin{document}

\newcommand{\sqvb}{\ensuremath{ \langle \!\langle 0 |} }
\newcommand{\sqvk}{\ensuremath{ | 0 \rangle \!\rangle } }
\newcommand{\sqvn}{\ensuremath{ \langle \! \langle 0 |  0 \rangle \! \rangle} }
\newcommand{\wh}{\ensuremath{\widehat}}
\newcommand{\be}{\begin{equation}}
\newcommand{\ee}{\end{equation}}
\newcommand{\bea}{\begin{eqnarray}}
\newcommand{\eea}{\end{eqnarray}}
\newcommand{\ra}{\ensuremath{\rangle}} 
\newcommand{\la}{\ensuremath{\langle}}
\newcommand{\rra}{\ensuremath{ \rangle \! \rangle }}
\newcommand{\lla}{\ensuremath{ \langle \! \langle }}
\newcommand{\str}{\rule[-.125cm]{0cm}{.5cm}}
\newcommand{\pr}{\ensuremath{^\prime}}
\newcommand{\ppr}{\ensuremath{^{\prime \prime}}}
\newcommand{\da}{\ensuremath{^\dag}}
\newcommand{\as}{^\ast}
\newcommand{\eps}{\ensuremath{\epsilon}}
\newcommand{\ve}{\ensuremath{\vec}}
\newcommand{\ka}{\kappa}
\newcommand{\non}{\ensuremath{\nonumber}}
\newcommand{\lf}{\ensuremath{\left}}
\newcommand{\rt}{\ensuremath{\right}}
\newcommand{\al}{\ensuremath{\alpha}}
\newcommand{\dfn}{\ensuremath{\equiv}}
\newcommand{\ga}{\ensuremath{\gamma}}
\newcommand{\ti}{\ensuremath{\tilde}}
\newcommand{\wti}{\ensuremath{\widetilde}}
\newcommand{\hs}{\ensuremath{\hspace*{.5cm}}}
\newcommand{\bet}{\ensuremath{\beta}}
\newcommand{\om}{\ensuremath{\omega}}

\newcommand{\cO}{\ensuremath{{\cal O}}}
\newcommand{\cS}{\ensuremath{{\cal S}}}
\newcommand{\cF}{\ensuremath{{\cal F}}}
\newcommand{\cX}{\ensuremath{{\cal X}}}
\newcommand{\cZ}{\ensuremath{{\cal Z}}}
\newcommand{\cG}{\ensuremath{{\cal G}}}
\newcommand{\cR}{\ensuremath{{\cal R}}}
\newcommand{\cV}{\ensuremath{{\cal V}}}
\newcommand{\cC}{\ensuremath{{\cal C}}}
\newcommand{\cP}{\ensuremath{{\cal P}}}
\newcommand{\pup}{\ensuremath{^{(p)}}}
\newcommand{\prpr}{\ensuremath{\prime \prime }}

\twocolumn[
\OSAJNLtitle{\bf 
Squeezing the Local Oscillator Does Not Improve Signal-to-Noise Ratio in Heterodyne Laser Radar
}
\OSAJNLauthor{
Mark A. Rubin and Sumanth Kaushik}
\OSAJNLaddress{Lincoln Laboratory\\ 
Massachusetts Institute of Technology\\  
244 Wood Street\\                         
Lexington, Massachusetts 02420-9185}
\OSAJNLemail{\{rubin,skaushik\}@LL.mit.edu}

\begin{center}
\begin{abstract*}
The signal-to-noise ratio for  heterodyne laser radar with a  coherent target-return beam and a squeezed
local-oscillator beam is  lower than that obtained using a coherent local oscillator,
regardless of the method employed to combine  the  beams at the detector.

\end{abstract*}
\end{center}

]
\maketitle\symbolfootnote[0]{This work was sponsored by the Air Force under Air Force Contract 
FA8721-05-C-0002.  Opinions, interpretations, conclusions, and 
recommendations are those of the authors and are not necessarily endorsed 
by the U.S. Government.}

Squeezed light holds promise for reducing noise in optical-detection applications below 
the level obtainable using  coherent light such as that  emitted by lasers.\cite{WallsMilburn1994} 
However, squeezing is  degraded by loss\cite{Caves1981}. 
In  laser radar applications, the loss in the target-return beam---the beam received by the radar
system after reflection from a target---is  severe.\cite{Kingston1995} 
Therefore squeezing is not useful in laser radar, at least not when applied to the target-return beam.

This still leaves open the possibility that squeezed light could be profitably employed as the local oscillator (LO) in a {\rm heterodyne}\/  laser radar. In such a system the target-return beam is combined  with a local-oscillator beam of a different frequency on a photosensitive detector, and the presence of a target is inferred from observation of oscillation of the detector response at the difference frequency of the two beams\cite{Kingston1995}.
A heterodyne laser radar system combining the target-return and LO beams on a single  detecting element has been proposed by 
Li {\em et al.}\/\cite{Lietal1999}. Their work has been criticized by Ralph\cite{Ralph2000} on the grounds that
the method they employ to combine the target-return beam and LO beam on the detector introduces sufficient noise
to cancel out any improvement in signal-to-noise ratio (SNR) due to squeezing.  However, one might envision employing other
methods of combining the beams  which do not add noise; e.g., using a Fabry-Perot etalon\cite{Hernandez1986} that reflects the 
LO frequency and transmits the target-return frequency.

(This approach is to be distinguished from a ``balanced''
heterodyne system in which the two beams are directed to {\em two}\/ detectors using a beam splitter. Quantum
noise in balanced heterodyne detection with  squeezed light has been examined by Yuen and Chan\cite{YuenChan1983} and 
by Annovazzi-Lodi {\em et al.}\/\cite{Annovazzi-Lodietal1992}.   We would not expect such
a system to  benefit from squeezing, since both beams must pass through the beam splitter and thus suffer
squeezing-destroying loss. In fact, a calculation of the SNR in  the balanced case 
gives the same result  as that obtained below in the present case, eq. (\ref{SNRb}), except for the  change of $\cos^2$\/ to $\sin^2$\/ due to the
$\pi/2$\/ phase change of the reflected light at the beam splitter. The detection of a Doppler beat signal using
a squeezed LO, as proposed by Li {\em et al.}\/\cite{Lietal1997}, is also an example of
a heterodyne measurement to which the considerations of the present paper apply.)

Here, we show that  a heterodyne detection scheme combining a coherent target-return beam and a squeezed LO beam on a single detector will 
fail to improve SNR regardless of the method used to combine the beams.

For target detection using the statistic $S$\/\cite{Helstrom1976}, 
\be
\mbox{SNR}=\left(\la\psi_1|\wh{S}|\psi_1\ra-\la\psi_0|\wh{S}|\psi_0\ra\right)^2/\mbox{Var}_0S,\label{SNRdef}
\ee
where $\la\psi_1|\wh{S}|\psi_1\ra$\/  
is the mean value of $S$\/ in  that quantum state, $|\psi_1\ra$\/, in which the target is  present, 
$\la\psi_0|\wh{S}|\psi_0\ra$\/  
is the mean value of $S$\/  when the target is absent, and 
$\mbox{Var}_0S$\/ is the variance of $S$\/ in the target-absent condition,
\be
\mbox{Var}_0S=\la \psi_0|\wh{S}^2|\psi_0\ra-\la \psi_0|\wh{S}|\psi_0\ra^2.\label{Var0formula}
\ee
In choosing pure quantum states to correspond to the target-present and target-absent conditions we are assuming 
the absence of additional non-quantum sources of noise, e.g., thermal noise, which would have to be treated using the density
operator formalism.\cite{GottfriedYan2003} 

For heterodyne detection,
\be
\wh{S}=\tau^{-1}\int_0^\tau dt \; \cos(\omega_Ht+\theta_H)\wh{I}(t), \label{Sopdef}
\ee
where $\wh{I}(t)$\/ is the quantum operator corresponding to the photoelectric current produced by the
detector at time $t$\/ and $\tau$\/ is the fixed time interval during which the target present/absent
decision is to be made.  That is,  $\wh{S}$\/ is the Fourier component of the photoelectric current at frequency $\omega_H$\/ and phase $\theta_H$\/.
We take $\omega_H$\/ to be equal to the difference
between the respective frequencies of the target-return and LO
beams,
\be
\omega_H=\omega_T-\omega_{LO}.\label{omegaorder}
\ee
(For simplicity we will always take  $\omega_T-\omega_{LO} >0$\/.) 
So $\wh{S}$\/ corresponds to detection of the beat frequency between the LO and the target-return beam.

For suitable broadband detectors,
the operator corresponding to the photoelectric current at time $t$\/  is\cite{Glauber1965}
\be
\wh{I}(t)=\kappa \wh{E}^{(-)}(t)\;\wh{E}^{(+)}(t),\label{IopdefP2}
\ee
where $\kappa$\/ is a  constant, $\wh{E}^{(-)}(t)=(\wh{E}^{(+)}(t))\da,$\/ and $\wh{E}^{(+)}(t)$\/ is the positive-frequency part of the time-dependent electric field operator at the
detector,  
%
\be
\wh{E}^{(+)}(t)=\sum_{k}i\left(\frac{\hbar \omega_{k}}{2 \varepsilon_0 V}\right)^{1/2}
\wh{a}_{k}\; \exp\left(-i\omega_{k}t\right).\label{EplusP2}
\ee
The mode frequencies are $\omega_k=ck$\/ 
 where the wavenumber $k$\/ runs over the values $k=2\pi n/V^{1/3}$\/, $n=1,2,\ldots.$\/
In writing  $\wh{E}^{(+)}(t)$\/ as in  (\ref{EplusP2}) we are assuming that the detector is only sensitive to a single direction of polarization (which is the
direction in which both the LO and target-return beam will be polarized) and that the optical system is such that, for
each frequency,  only
a single spatial mode need be considered (that mode with wave vector normal to the detector surface). 
The annihilation operators $\wh{a}_k$\/ satisfy the usual commutation relations,
\be
[\wh{a}_k,\wh{a}_l]=[\wh{a}\da_k,\wh{a}\da_l]=0, \hspace*{5mm}[\wh{a}_k,\wh{a}\da_l]=\delta_{kl}.\label{commutation}
\ee
Using (\ref{Sopdef})-(\ref{EplusP2}) 
we obtain, in the limit $\tau \rightarrow \infty$\/,
\bea
\wh{S}&=&\frac{\kappa\hbar}{4 \varepsilon_0 V}\;
\sum_{l, k \mbox{ \footnotesize s.t. } |\omega_{l}-\omega_{k}|=\omega_H}
(\omega_{l}\omega_{k})^{1/2}\nonumber\\
&&\wh{a}\da_{l}\wh{a}_{k}
\;\exp\left(-i\varepsilon(\omega_{l}-\omega_{k})\theta_H\right) ,\label{SopP2}
\eea
where 
\be
\varepsilon(x)=\mbox{ sign of }x. \label{signfuncdef}
\ee

In the target-absent state, all modes but the LO are in the vacuum state:
\be
|\psi_0\ra=|\alpha,\xi\ra _{k_{LO}} \prod_{k\neq k_{LO}} |0\ra_{k}.\label{psi0P2}
\ee
Here $|\alpha,\xi\ra _{k_{LO}}$\/ is the squeezed local-oscillator-frequency ($\omega_{LO}$\/) mode parameterized by
complex numbers $\alpha$\/ and $\xi$\/\cite{GerryKnight2005}.  
In the target-present case
an additional mode is in a nonvacuum state, specifically the coherent state $|\beta\ra_{k_T}$\/ at the target-return frequency $\omega_{T}$\/:
\be
|\psi_1\ra=|\beta\ra_{k_T}|\alpha,\xi\ra_{k_{LO}} \prod_{k \neq k_T,k_{LO}} |0\ra_{k}.\label{psi1P2}
\ee

Using (\ref{omegaorder}), (\ref{SopP2})-(\ref{psi1P2}) 
and the relations\cite{GerryKnight2005}
\be
\wh{a}_k|0\ra_k=\mbox{}_k\la 0|\wh{a}\da_k=0,\label{killers}
\ee
\be
\mbox{}_{k_{T}}\la \beta |\wh{a}_{k_{T}}|\beta\ra_{k_{T}}=\beta,\hspace*{5mm}
\mbox{}_{k_{T}}\la \beta |\wh{a}\da_{k_{T}}|\beta\ra_{k_{T}}=\beta^\ast,\label{cohmatelts}
\ee
and
\bea
\mbox{}_{k_{LO}}\la \alpha,\xi |\wh{a}_{k_{LO}}|\alpha,\xi\ra_{k_{LO}}&=&\alpha,\nonumber\\
\mbox{}_{k_{LO}}\la \alpha,\xi |\wh{a}\da_{k_{LO}}|\alpha,\xi\ra_{k_{LO}}&=&\alpha^\ast,\label{sqmatelts}
\eea
we find that
\be
\la \psi_0|\wh{S}|\psi_0 \ra=0,\label{meanS0}
\ee
since the only possible nonzero term, $\wh{a}\da_{k_{LO}}\wh{a}_{k_{LO}}$\/, is forbidden by the restriction on
the summation  in (\ref{SopP2}),
and
\bea
\la \psi_1|\wh{S}|\psi_1 \ra&=&\frac{\kappa \hbar}{2\varepsilon_0 V}\;\left(\omega_{T}\omega_{LO}\right)^{1/2}\nonumber\\
&&|\alpha||\beta|\cos(\theta_T-\theta_{LO}+\theta_H),\label{meanS1}
\eea
where
\be
\theta_T=\arg{\beta},\hspace*{5mm}\theta_{LO}=\arg{\alpha}.\label{thetadefs}
\ee

Using (\ref{commutation})-(\ref{psi0P2}) and (\ref{killers}),
$$
\la \psi_0|\wh{S}^2|\psi_0 \ra=\left(\frac{\kappa\hbar}{4 \varepsilon_0 V}\right)^2\hspace*{5in}
$$
$$
\sum_{k \mbox{ \footnotesize s.t. }|\omega_{LO}-\omega_{k}|=\omega_H}\;\;
\sum_{l \mbox{ \footnotesize s.t. }|\omega_{l}-\omega_{LO}|=\omega_H}\;\; 
\omega_{LO}\left(\omega_{k}\omega_{l}\right)^{1/2}\hspace*{5in}
$$
\bea
&&
\la\psi_0|\wh{a}\da_{k_{LO}}\wh{a}_{k}\wh{a}\da_{l}\wh{a}_{k_{LO}}|\psi_0\ra\nonumber\\
&&\exp\left(-i[\varepsilon(\omega_{LO}-\omega_{k})+\varepsilon(\omega_{l}-\omega_{LO})]\theta_H\right).\label{meanSsqaP2}
\eea
Neither $k$\/ nor $l$\/ can be equal to $k_{LO}$\/, due to the restrictions in the summations in (\ref{meanSsqaP2}). 
If $k\neq l$\/ then $\wh{a}_{k}$\/ and $\wh{a}\da_{l}$\/
commute, yielding zero since the non-$LO$\/ modes are in the vacuum state. So the only surviving terms are
those for which $k=l$\/. Using (\ref{commutation}),
\be
\la \psi_0|\wh{S}^2|\psi_0 \ra=\left(\frac{\kappa\hbar}{4 \varepsilon_0 V}\right)^2
\sum_{k \mbox{ \footnotesize s.t. }|\omega_{LO}-\omega_{k}|=\omega_H}\omega_{LO}\omega_{k}
\bar{n}_{LO},\label{meanSsqbP2}
\ee
where
\be
\bar{n}_{LO}=\mbox{}_{k_{LO}}\la \alpha,\xi|\wh{a}\da_{k_{LO}}\wh{a}_{k_{LO}}|\alpha,\xi\ra_{k_{LO}}.\label{nLObarP2}
\ee
Using (\ref{Var0formula}), (\ref{meanS0}) and (\ref{meanSsqbP2}),
\be
\mbox{Var}_0S=\left(\frac{\kappa\hbar}{4 \varepsilon_0 V}\right)^2
\sum_{k \mbox{ \footnotesize s.t. }|\omega_{LO}-\omega_{k}|=\omega_H}\omega_{LO}\omega_{k}
\bar{n}_{LO}.\label{Var0bP2}
\ee
The contribution to (\ref{Var0bP2}) from the term $\omega_k=\omega_{LO}-\omega_H$\/ is termed the ``image band''
contribution\cite{Haus2000}. 

In practice $\omega_H \ll \omega_T,\omega_{LO}$\/,  
so we can take
\be
\omega_T\approx\omega_{LO}\equiv\omega.\label{omega}
\ee
Using (\ref{omega}), (\ref{meanS1}) and (\ref{Var0bP2}) become
\be
\la \psi_1|\wh{S}|\psi_1 \ra=\frac{\kappa \hbar\omega}{2\varepsilon_0 V}
|\alpha||\beta|\cos(\theta_T-\theta_{LO}+\theta_H),\label{meanS1bP2}
\ee
\bea
\mbox{Var}_0S&=&2\left(\frac{\kappa\hbar\omega}{4 \varepsilon_0 V}\right)^2
\bar{n}_{LO}.\label{Var0cP2}
\eea
Using (\ref{meanS0}), (\ref{meanS1bP2}) and (\ref{Var0cP2}), the signal-to-noise ratio (\ref{SNRdef}) is 
$$
\mbox{SNR}=\frac{2|\alpha|^2|\beta|^2 \cos^2(\theta_T-\theta_{LO}+\theta_H)}{\bar{n}_{LO}}\hspace*{5in}
$$
\be
=2\left(1-\frac{\sinh^2(r)}{\bar{n}_{LO}}\right)\bar{n}_T \cos^2(\theta_T-\theta_{LO}+\theta_H)\label{SNRb}
\ee
using the relations\cite{GerryKnight2005}
\be
|\beta|^2=\bar{n}_{k_T}=\mbox{}_{k_T}\la \beta|\wh{a}\da_{k_T}\wh{a}_{k_T}|\beta\ra_{k_T},\label{nTP2}
\ee
\be
\bar{n}_{LO}=|\alpha|^2+\sinh^2(r).\label{nLOalphaconstraint}
\ee
The  parameter $r=|\xi|$\/ 
is termed the ``squeezing parameter.'' The value $r=0$\/ corresponds to no squeezing (coherent state).
From (\ref{SNRb}) it is clear that squeezing the LO mode---i.e., letting the LO be in
a state with $r > 0$\/---can only reduce the signal-to-noise ratio.

This result is consistent with the observation by Yuen and Chan\cite{YuenChan1983}, in the context of balanced detection, that  while ``quantum noise is frequently supposed to
arise from local-oscillator (LO) shot noise \ldots it actually arises from the signal quantum fluctuation.''
The reasonable but incorrect expectation that squeezing the LO will improve SNR arises from the fact that
the variance of the zero-frequency signal, i.e., the 
time-averaged photoelectric current corresponding
to the operator
\be
\wh{S}\pr=\tau^{-1}\int_0^\tau dt \; \wh{I}(t)=\frac{\kappa\hbar\omega_{k}}{2 \varepsilon_0 V}\;\wh{a}\da_{k}\wh{a}_{k} \label{Spropdef}
\ee
(the second equalty holding in the limit $\tau \rightarrow \infty$\/),  does change with squeezing. 
In the target-absent state,
\be
\mbox{Var}_0 S\pr=\la\psi_0|\wh{S}^{\prime 2}|\psi_0\ra-\la\psi_0|\wh{S}^{\prime }|\psi_0\ra^2,\label{VaroSprdef}
\ee
which, for $\tau \rightarrow \infty$\/ has the  value
\be
\mbox{Var}_0 S\pr=\left(\frac{\kappa\hbar\omega_{LO}}{2\varepsilon V}\right)^2\mbox{var}\;n_{LO},\label{Var0Spr}
\ee
where
\bea
\mbox{var}\;n_{LO}&=&\left[\rule[-.3cm]{0cm}{.6cm}\mbox{}_{k_{LO}}\la \alpha,\xi|(\wh{a}\da_{k_{LO}}\wh{a}_{k_{LO}})^2|\alpha,\xi\ra_{k_{LO}}\right.\nonumber\\
&-&\left.\left(\mbox{}_{k_{LO}}\la \alpha,\xi|\wh{a}\da_{k_{LO}}\wh{a}_{k_{LO}}|\alpha,\xi\ra_{k_{LO}}\right)^2\right].\label{varnLOdef}
\eea
For suitable choice of the phase of $\xi$\/, (\ref{varnLOdef}) can indeed be lower than $\bar{n}_{LO}$\/, the value it takes in a 
coherent state.\cite{GerryKnight2005}

However, statistical decision theory\cite{Helstrom1976} indicates that if a quantity $S$\/ computed from measurements made  by a detector is used as the decision criterion in a target-detection task, then it is the variance of {\em that
same quantity $S$\/}\/  that is relevant in evaluating the suitability of $S$\/ for the task. In heterodyne radar the computed quantitiy $S$\/ is the Fourier component of the instantaneous photocurrent induced at the detector by the combined target-return and LO beams\cite{Kingston1995}. The operator corresponding to the instantaneous detector response is $\wh{I}(t)$\/, so  the operator corresponding to the required Fourier component is $\wh{S}$\/ as defined in (\ref{Sopdef}). 
It is thus the variance of $\wh{S}$\/, not that  of 
$\wh{S}\pr$\/, which must be used for computing SNR.



(The general expression for the signal operator (\ref{Sopdef})  for 
$\tau$\/
not necessarily infinite, $\omega_H$\/ not necessarily equal to $|\omega_k-\omega_l|$\/ for any $k,l$\/, is
$$
\wh{S}=\frac{\kappa\hbar}{2 \varepsilon_0 V}\;\sum_{l,k} (\omega_{l}\omega_{k})^{1/2} \wh{a}\da_{l}\wh{a}_{k}\exp\left(-i\varepsilon(\omega_{l}-\omega_{k})\theta_H\right)
$$
\vspace*{-.375cm}
\bea
\cdot\frac{1}{2i\tau}\left\{\frac{\exp\left(i\theta_H\right)}{\omega_l-\omega_k+\omega_H}
\left[\exp\left(i(\omega_l-\omega_k+\omega_H)\tau\right)-1\right]\right.&&\nonumber\\
\hspace{6mm}\left.+\frac{\exp\left(-i\theta_H\right)}{\omega_l-\omega_k-\omega_H}
\left[\exp\left(i(\omega_l-\omega_k-\omega_H)\tau\right)-1\right]\right\}.&&\label{Gkldef}
\eea
This reduces to (\ref{SopP2}), (\ref{Spropdef})  
for the appropriate limiting values of $\tau$\/, $\omega_H$\/ and $\theta_H$\/.)

\mbox{}

M. A. R. thanks Jonathan Ashcom and  Jae Kyung for a helpful discussion on mixing efficiency.


\begin{thebibliography}{99}
\bibitem{WallsMilburn1994} D.~F.~Walls and G.~J.~Milburn, {\em Quantum Optics}\/ (Springer, Berlin, 1994).
\bibitem{Caves1981} C.~M.~Caves,  Phys. Rev. D {\bf23}, 1693 (1981).
\bibitem{Kingston1995}  R.~H.~Kingston, {\em Optical Sources, Detectors and Systems: Fundamentals and Applications}\/, (Academic Press, San Diego, 1995).
\bibitem{Lietal1999} Y.-q.~Li, D.~Guzun, and M.~Xiao, Phys. Rev. Lett. {\bf 82}, 5225 (1999).
\bibitem{Ralph2000} T.~C.~ Ralph, Phys. Rev. Lett. {\bf 85}, 677 (2000).
\bibitem{Hernandez1986}  G.~Hernandez, {\em Fabry-Perot Interferometers}\/ (Cambridge, 1986).
\bibitem{YuenChan1983} H.~P.~Yuen and  and V.~W.~S.~Chan, Opt. Lett. {\bf 8}, 177 (1983).
\bibitem{Annovazzi-Lodietal1992}V.~Annovazzi-Lodi, S.~Donati and S.~Merlo,  Opt. Quan.  Electron. {\bf 24},  258 (1992).
\bibitem{Lietal1997}  Y.-q.~Li, P.~Lynam, M.~Xiao, and P.~J.~Edwards, Phys. 
	Rev. Lett.  {\bf 78}, 3105 (1997).
\bibitem{Helstrom1976} C.~W.~Helstrom, {\em Quantum Detection and Estimation Theory}\/ (Academic, New York, 1976).
\bibitem{GottfriedYan2003}  K.~Gottfreid and T.-M.~Yan, {\em Quantum Mechanics: Fundamentals,}\/ 2nd. ed.  (Springer, New York, 2003).
\bibitem{Glauber1965}  R.~J.~Glauber, ``Optical coherence and photon statistics,'' in
	{\em Quantum Optics and Electronics},  C.~DeWitt, A.~Blandin, and C.~Cohen-Tannoudji, eds. (Gordon and Breach, New York, 1965).
\bibitem{GerryKnight2005}  C.~G.~Gerry and  P.~L.~Knight, {\em Introductory Quantum Optics}\/ (Cambridge, 2005).
\bibitem{Haus2000}  H.~A.~Haus, {\em Electromagnetic Noise and Quantum Optical Measurements}\/ (Springer, Berlin, 2000).
\end{thebibliography}
\end{document}